\documentclass[preprint,12pt,5p,twocolumn]{elsarticle}
\biboptions{sort&compress} 

\usepackage{amssymb}
\usepackage{amsmath, cuted} 
\AfterEndEnvironment{strip}{\leavevmode}
\usepackage{graphicx}
\usepackage{caption}
\usepackage{subcaption}
\usepackage{multirow}
\usepackage{multicol}
\usepackage{hyperref}
\usepackage{natbib}
\usepackage{url}
\usepackage{xcolor}
\usepackage{soul}
\usepackage{afterpage}
\usepackage{fancyhdr}

\journal{Journal of Network and Computer Applications}

\fancypagestyle{firstpage}{%
 \fancyhf{}%
 \fancyhead[L]{\centering\tiny{This is the authors' version of the manuscript accepted in the Journal of Network and Computer Applications (2024).\\Final publication can be found in doi: https://doi.org/10.1016/j.jnca.2024.103939.\\This manuscript version is made available under the CC-BY-NC-ND 4.0 license https://creativecommons.org/license/by-nc-nd/4.0.}}
}

\begin{document}

\begin{frontmatter}



\title{Virtual Reality Traffic Prioritization for Wi-Fi Quality of Service Improvement using Machine Learning Classification Techniques}


\author[inst1]{Seyedeh Soheila Shaabanzadeh\corref{cor1}}
\cortext[cor1]{Corresponding author}

\affiliation[inst1]{organization={Signal Theory and Communications},
            addressline={Universitat Politècnica de Catalunya (UPC), Carrer de Jordi Girona, 1-3}, 
            postcode={08034}, 
            state={Barcelona},
            countof Service (QoS) with giving priority to VR users in the access point (AP) and efficiently handle VRry={Spain}}

\author[inst2]{Marc Carrascosa-Zamacois}

\author[inst1]{Juan Sánchez-González}

\author[inst2]{Costas Michaelides}
\affiliation[inst2]{organization={Information and Communications Technologies},
            addressline={Universitat Pompeu Fabra (UPF), Roc Boronat, 138}, 
            postcode={08018}, 
            state={Barcelona},
            country={Spain}}


\author[inst2]{Boris Bellalta}


\begin{abstract}
The increase in the demand for eXtended Reality (XR)/Virtual Reality (VR) services in the recent years, poses a great challenge for Wi-Fi networks to maintain the strict latency requirements. In VR over Wi-Fi, latency is a significant issue. In fact, VR users expect instantaneous responses to their interactions, and any noticeable delay can disrupt user experience. Such disruptions can cause motion sickness, and users might end up quitting the service. Differentiating interactive VR traffic from Non-VR traffic within a Wi-Fi network can aim to decrease latency for VR users and improve Wi-Fi Quality of Service (QoS) with giving priority to VR users in the access point (AP) and efficiently handle VR traffic. In this paper, we propose a machine learning-based approach for identifying interactive VR traffic in a Cloud-Edge VR scenario. The correlation between downlink and uplink is crucial in our study. First, we extract features from single-user traffic characteristics and then, we compare six common classification techniques (i.e., Logistic Regression, Support Vector Machines, k-Nearest Neighbors, Decision Trees, Random Forest, and Naive Bayes). For each classifier, a process of hyperparameter tuning and feature selection, namely permutation importance is applied. The model created is evaluated using datasets generated by different VR applications, including both single and multi-user cases. Then, a Wi-Fi network simulator is used to analyze the VR traffic identification and prioritization QoS improvements. Our simulation results show that we successfully reduce VR traffic delays by a factor of 4.2x compared to scenarios without prioritization, while incurring only a 2.3x increase in delay for background (BG) traffic related to Non-VR services.
\end{abstract}




\end{frontmatter}


\thispagestyle{firstpage}  

\section{Introduction}
\label{sec:introduction}
eXtended Reality (XR) represents the convergence of physical and virtual areas, creating immersive environments through technologies like Augmented Reality (AR), Virtual Reality (VR), and Mixed Reality (MR). XR’s profound interactivity and immersion offer captivating virtual experiences to users worldwide \cite{wifialliance}. The XR market is expected for significant growth, with projection indicating an increase from USD 105.58 billion in 2023 to USD 472.39 billion by 2028, at a robust Compound Annual Growth Rate (CAGR) of 34.94\% during the forecast period \cite{reportlinker}.

Wi-Fi can play a key role in facilitating VR experiences \cite{wifialliance}. A common setup is playing a video game on a powerful desktop, and using a Wi-Fi network to deliver it to the clients, usually Head Mounted Devices (HMDs), or simply VR headsets. In the downlink (DL), the desktop transmits VR game's video frames to the VR headset, allowing the user to experience the visual content. Simultaneously, in the uplink (UL), the motion data from the VR headset is sent back to the desktop. This feedback loop enables the desktop to generate the next VR video frame based on the user's latest viewport \cite{9013507, michaelides2023wi}. Through the paper, we will refer to this scenario as VR edge streaming over Wi-Fi. 

Latency is a significant issue in VR, both in wired and wireless systems \cite{9013507}. In VR services, a too high latency may cause that the gap between the VR video game and what the user sees becomes excessively large. This gap can lead to motion sickness and degrade the perceived user quality of experience. The inconsistency does not appear only because of transmission delay, but jitter of feedback in uplink as well \cite{9013507}. Wi-Fi QoS management can significantly enhance user experiences by reducing latency and jitter in real-time applications like gaming, preventing disruptions in immersive experiences, and minimizing lag during access to cloud and edge services \cite{wifiallianceqos}. 

The increasing demand for high data transmission has led to extensive research in creating new mechanisms to identify and classify network traffic. When these mechanisms are not managed effectively, it can negatively impact important network operational functions like Fault Tolerance, Traffic Engineering, Quality of Service (QoS) provisioning, and Dynamic Access Control. In essence, a precise traffic identification procedure allows the efficient management of existing network resources, thereby permitting more accurate and robust resource allocation schemes  \cite{dias2019innovative}. 

Traffic classification methods are typically categorized into three main groups, namely port-based, payload-based known as Deep Packet Inspection (DPI), and Machine Learning (ML)-based techniques \cite{dias2019innovative}. 
ML-based methodologies present a viable solution to circumvent certain constraints inherent in port-based and payload-based approaches. Specifically, ML techniques demonstrate the capacity to classify network traffic by leveraging traffic statistics that are independent of application protocols. These statistics encompass metrics such as flow duration, packet length variance, maximum and minimum segment sizes, window size, round trip time, and packet inter-arrival time. 
In light of the importance of network traffic classification and the need to identify VR traffic, the novelty of this paper is the proposal of an interactive VR ML traffic classification model in the edge streaming VR scenario. The proposed model evaluates the use of some common classifiers, fed with the features previously extracted. The model is evaluated by using VR traffic traces coming from a multi-user VR scenario, and using single-user traces from a VR framework not included in the training traces. Finally, a Wi-Fi network environment is simulated in order to  illustrate how the model can be used to detect VR traffic, and give it higher priority to improve its QoS. In particular, the key contributions of this study are as follows: 

\begin{itemize}
    \item A binary classification ML model is proposed to identify interactive VR traffic. In this model, a heterogeneous traffic dataset is used with statistical features computed from single-user traffic characteristics over certain duration.

    \item The collected packet traces, the input files for the classification models (in CSV format) used for both training and test phases, along with the corresponding Python scripts, are available on Zenodo.

    \item  The proposed classification model is evaluated on packet traces from a multi-user experimental setup involving three users, and using single-user traces from a VR framework not included in the training traces. In addition, Wi-Fi Quality of Service (QoS) is enhanced by reducing VR traffic delay through prioritizing VR traffic. The proposed prioritization methodology has been evaluated by means of a Wi-Fi simulator.
\end{itemize}
The rest of the paper is organized as follows. Section 2 provides an overview of the existing literature. Section 3 describes the experimental design, including dataset creation, and traffic classification methods. The obtained results in both feature extraction and traffic classification are presented in Section 4, while the testing process of the model and the evaluation of VR traffic identification in a Wi-Fi environment simulation are discussed in Section 5. Finally, conclusion and future work are drawn in Section 6.


\section{Related Work} 
\label{secRelatedWork}

In the context of network traffic identification, the dynamic nature of ports and encryption techniques in both port-based and DPI methods has led to the increased prominence of machine learning-based approaches as widely adopted technologies for traffic identification. \cite{gibert2020rise, d2021network}. In recent years, scholars have made research into the domain of ML-based traffic identification.  Researchers in \cite{zhang2023video}, identify cloud game video traffic by proposing a novel approach for extracting distinctive features from video scenes to distinguish traffic patterns and then an adaptive distribution distance-based feature selection (ADDFS) technique for feature selection. In addition to conventional statistical features derived from standard network attributes (e.g. packet payload size, packet inter-arrival time), the authors defined the 'peak point' as the maximum data transmission in a given period for the feature extraction process. This innovation aimed to enhance the video identification process, reducing unnecessary costs, time overhead, and potential negative impacts on the identification model. 

In the paper \cite{10154417}, researchers focused on classifying Cloud Gaming (CG) traffic to optimize the use of a low-latency queue within the L4S (Low Latency, Low Loss, Scalable Throughput) architecture. The objective is to minimize latency fluctuations that could negatively impact Quality of Service (QoS) at the network edge. Three classification models (Thresholds, DT, RF) were developed and evaluated to distinguish CG traffic from other high-bitrate UDP-based applications. These models were constructed using 12 features derived from network flow data, including packet sizes and Inter-Arrival Times (IAT). Additionally, the study introduced a fully functional implementation of the classifier in the form of a micro-service architecture, suitable for deployment as Virtual Network Functions (VNFs) following the Network Function Virtualization (NFV) paradigm. 

In \cite{8108090, yuan2010svm}, researchers conducted model tuning in SVM and evaluated various SVM kernel functions, including linear, polynomial, sigmoid, and radial kernels, to classify Internet traffic categories such as database, mail, www, multimedia, game, and service. The results indicated that the Radial Kernel exhibited superior performance compared to the other kernels. The researchers employed a sequential forward feature selection algorithm to improve accuracy by reducing irrelevant and redundant features. Notably, in \cite{8108090}, the authors compared the results of SVM with unsupervised K-means clustering. The paper \cite{jenefa2018upgraded} compared the accuracy of C5.0 with C4.5, SVM, and NB learning algorithms. Extracted features comprised packet rate, data rate, and inter-arrival time statistics (minimum, mean, maximum, and standard deviation). Using a private dataset of 17 applications (e.g. FTP, HTTP, Skype, Game, BitTorrent) collected in their labs, the experimental evaluation demonstrated C5.0's superior performance.

In the paper \cite{dias2019innovative}, the authors introduced a ML framework based on NB designed for real-time application classification, particularly focusing on video network traffic. The framework was trained using 13 statistical features, encompassing packet arrival time, the average of decimal values, and the average value of IP datagram length. Feature extraction involved LAN frame length and IPv4 header fields, utilizing a feature extractor script in a three-packet window. This calculated values like Time between Packets, Mean and Variance of Time between Packets, IP Total Length, Time to Live, and Protocol field. They collected the dataset in their labs, comprising YouTube and Netflix videos, along with files downloaded. The classification module effectively distinguished between Netflix streaming video, YouTube streaming video, and background traffic, particularly involving two file downloads.

In the paper \cite{sun2018internet}, the authors tackled the challenge of reducing the computational resources required by the traditional SVM learning classifier in the context of internet traffic classification. They introduced Incremental SVM (ISVM), a concept aimed at decreasing the high training costs associated with memory and CPU usage. Additionally, the authors put forth the Authenticator ISVM (AISVM) framework, leveraging valuable information from prior training datasets. Comparative experiments with NB and NBKDE (Naive Bayes algorithm with kernel density estimation) learning algorithms revealed lower accuracy than SVM. The study demonstrated that the proposed frameworks (ISVM and AISVM) yielded higher accuracy results while utilizing fewer computational resources than SVM. Note that features included simple statistics about packet length, inter-packet time, and information derived from traffic flows. The study \cite{cao2020improved}, aimed to boost the accuracy of a traffic classification model based on SVM. It introduced two modules: one for feature selection and another for optimizing the SVM algorithm. The feature module used an innovative algorithm to extract the most representative features, while the classifier module employed an Improved Grid Search algorithm to enhance parameter selection. Evaluation on the Moore dataset showed high accuracy compared to SVM, NB, and kNN algorithms in a smaller feature space. 

The paper \cite{khatouni2019integrating}, focused on the identification of Social Media, Audio, and Video applications, particularly when they are encrypted. A machine learning-based approach was assessed using a comprehensive dataset gathered from four distinct networks and generated via four different off-the-shelf traffic flow exporters. This dataset encompassed a diverse array of services, including Web Browsing, Email, Chat, Streaming, File transfer, VoIP, and P2P. The evaluation of the proposed system primarily centers on its accuracy in classifying the aforementioned applications. Within this investigation, four distinct feature sets derived from various off-the-shelf network traffic flow exporters (Tstat, Tranalyzer, SiLK, and Argus) were employed for the classification task using DT. While some features were common across the four exporters, each computed them differently (e.g., average Round Trip Time). Furthermore, certain features were exclusive to specific exporters (e.g., IP Time To Live change count exists solely within the Tranalyzer feature set).

The paper \cite{dong2021multi} introduced an improved SVM approach called cost-sensitive SVM (CMSVM) to address accuracy, computational time, and data imbalance issues. CMSVM incorporates an active learning technique to dynamically assign weights to specific applications. Evaluation using \text{MOORE\_SET} and \text{NOC\_SET} datasets for classifying network flows into service groups demonstrated that the proposed solution is more effective than the traditional SVM classifier in terms of accuracy and addressing data imbalance. The paper \cite{afuwape2021performance} examined the effectiveness of classifying VPN and non-VPN network traffic using ensemble classifiers, focusing on precision, recall, and F1-score. Their experiments, using the ISCX dataset, revealed that Gradient Boosting (GB) and RF ensemble classifiers outperformed single classifiers like DT, Multi-Layer Perceptron (MLP), and kNN in terms of accuracy. 
In another work, The paper \cite{ganesan2021sdn} introduced a machine learning scheduling framework for prioritizing network traffic in an IoT environment based on QoS requirements. The study compared the performance of seven supervised learning algorithms, including RF, kNN, MLP, NB, LR, and SVM. Using the UNSW dataset with 21 IoT/non-IoT devices, RF achieved the highest accuracy results.

This paper's primary focus is the application of ML classifiers for identifying interactive VR traffic, a topic not thoroughly explored in prior research. VR traffic comprises both UL (user tracking) and DL (video batches) data, and it exhibits a distinct relationship between these two components. The interactive nature of VR traffic ---in contrast to video streaming alone--- makes it essential to differentiate from non-VR traffic, necessitating high-priority QoS profiles in Wi-Fi networks to ensure optimal performance and user experience.


\section{Experimental Design} \label{secmethod}

In this section, we describe the process followed for the generation of the datasets. We also present the definition and selection of the features used for classification, the design of the proposed classifiers and the aspects related to the tuning the hyperparameters of the considered classifiers\footnote{The raw packet traces, the input files and Python scripts for both the feature extraction and traffic classification are available on Zenodo \cite{shaabanzadeh2023virtual}. We also reserve some special traces for the final test}. 


\subsection{Dataset Creation} \label{subsecdata}

The dataset that is used in this paper consists on two different traffic datasets, one for VR traffic and one for Non-VR traffic. In our case, the first VR traffic dataset is related to a range of VR applications, including SteamVR Home, Half Life: Alyx, Budget Cuts, and a Custom Game developed in Unity called Alteration Hunting, in different configurations involving three distinct frame rates of 60, 90 and 120 frames per second (fps), as well as three different bit rates of 40, 50 and 100 Mbps. The second dataset representing Non-VR traffic focuses on commonly used application types such as non-game videos (i.e., multimedia streaming services like Youtube, Netflix), and online meetings (i.e., Zoom, Google Meet). This section provides an overview of the network environment in which the dataset was generated, outlines the data collection process, and presents the analysis of traffic data traces for both VR and Non-VR traffic.
\begin{enumerate}
    \item \textbf{Network Setup:} the network setup consists of a Desktop computer, access point (AP) that supports Wi-Fi 6 \cite{9442429, bellalta2016ieee, khorov2018tutorial}, a Head Mounted Display Meta Quest 2 and a laptop, as shown in Fig. \ref{fig:equipment}. The desktop computer is connected to the Access Point (AP) with an ethernet cable and the clients are connected to the AP via Wi-Fi. The equipment used is also illustrated in Table \ref{tab:table1}.

    \begin{table}[t]
      \caption{Equipment.}
      \centering
          \scriptsize 
          \setlength{\tabcolsep} {4pt} 
          \resizebox{\columnwidth}{!}{
              \begin{tabular}{|c|c|c|p{1cm}|}
                \hline
                \multirow{5}{*}{Desktop} & \multirow{2}{*}{OS} & Windows 10 (Unity),\\
                &  & Ubuntu 22.04 LTS (Steam) \\
                \cline{2-3}
                & CPU & 12th Gen Intel Core i5 \\
                \cline{2-3}
                & GPU & NVIDIA GeForce RTX 3080 \\
                \cline{2-3}
                & RAM & 2 x Kingston 16GB DDR5 \\
                \hline
                \multirow{3}{*}{Laptop} & OS & Ubuntu 22.04 LTS \\
                \cline{2-3} 
                & CPU & 11th Gen Intel Core i7 \\
                \cline{2-3} 
                & RAM & 16GB \\
                \hline
                HMD & Model & Meta Quest 2 \\
                \hline
                AP & Model & RT-AX58U \\
                \hline
              \end{tabular}
          }
      \label{tab:table1}
    \end{table}
    
    For VR, we have used two platforms to play games: Steam and Unity. In the Steam platform, the Desktop computer and HMD in Fig. \ref{fig:figure1} are used as a server and a client, respectively for streaming. We install the SteamVR\footnote{https://www.steamvr.com} platform, and the Air light VR (ALVR)\footnote{https://github.com/alvr-org/ALVR} server on the Desktop in order to allow users to play PC-based VR games on their HMD via a Wi-Fi connection. We also connect HMD as a client using the ALVR application. So, Playing games here is done on the HMD while the Desktop handle rendering and processing, streaming DL data (i.e. game content like video, audio and haptics) to the HMD and UL data (i.e. pose tracking, controller input, and performance metrics) from the HMD back to the Desktop for processing. In the Unity platform, the Desktop computer and laptop in Fig.1 are used as a server and a client, respectively for streaming. In order to develop a custom VR game, first the VR game is created and integrated in Unity, enabling VR support for the laptop. DL and UL traffic in Unity is similar to ALVR.
    
    For Non-VR, we just made use of the laptop connected to the AP linked to the internet in the Fig.\ref{fig:setup}. For the streaming process of multimedia services, we played the content on the laptop. The streaming operation involves sending requests for video and audio content as UL data and in response, downloading video content from multimedia servers as DL data. For the streaming process of online meetings, the process involves transmitting audio and video data as UL information while receiving incoming video and audio streams as DL data from the conference servers.
    
    \begin{figure}[t]
     \centering
     \begin{subfigure}[b]{0.5\textwidth}
         \centering
         \includegraphics[width=1\linewidth]{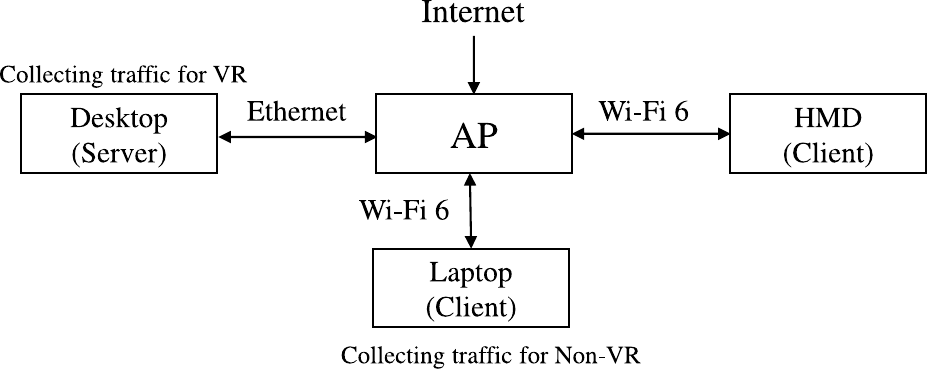} 
         \caption{}
         \label{fig:setup}
     \end{subfigure}
     \hfill
     \begin{subfigure}[b]{0.35\textwidth}
         \centering
         \includegraphics[width=1\linewidth]{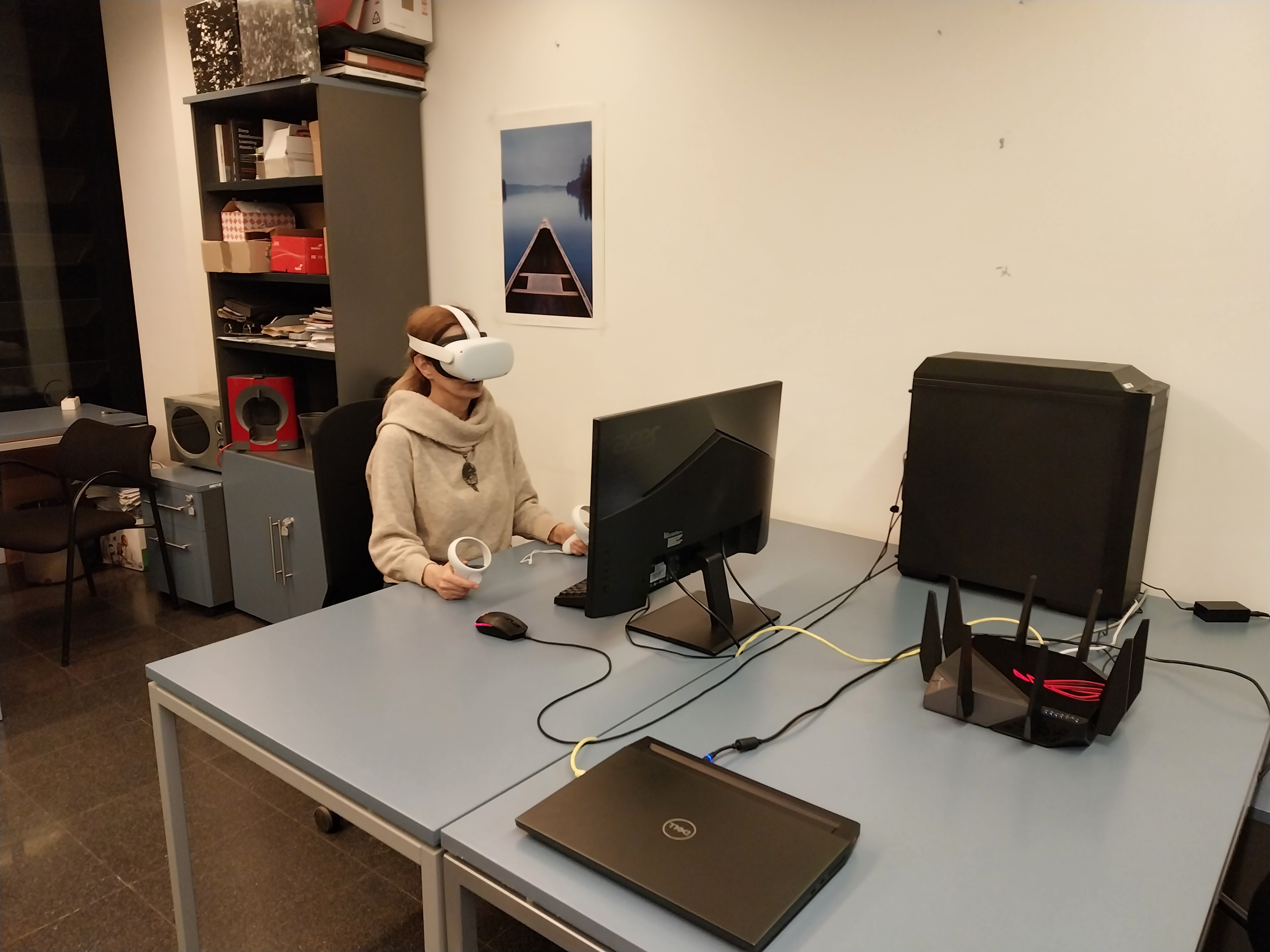}
         \caption{}
         \label{fig:equipment}
     \end{subfigure}
     \hfill
        \caption{The network setup for data collection (a) VR and Non-VR (b) Photo during VR experiment.}
        \label{fig:figure1}
    \end{figure}
    
    \item \textbf{Data Collection:} this section presents an overview of our dataset and the process by which we generated it. We collected VR and Non-VR traffic data in Desktop and laptop, respectively, as shown in Fig.1. In our study, we utilized Wireshark, a network protocol analyzer, to gather raw traffic data and to save as .pcap files. The collected traffic data are organized into distinct network flows based on their five-tuple information, comprising \{source IP address, source port number, destination IP address, destination port number, and transport protocol (i.e., UDP or TCP)\}. During the data collection process, we took measurements to minimize interference from non-targeted traffic by shutting down other applications. To provide a more clear understanding of the traffic traces collected within the VR and Non-VR contents, we briefly outline the traffic traces in the following:
    \begin{figure*}[t]
          \centering
              \includegraphics[width=1\linewidth]{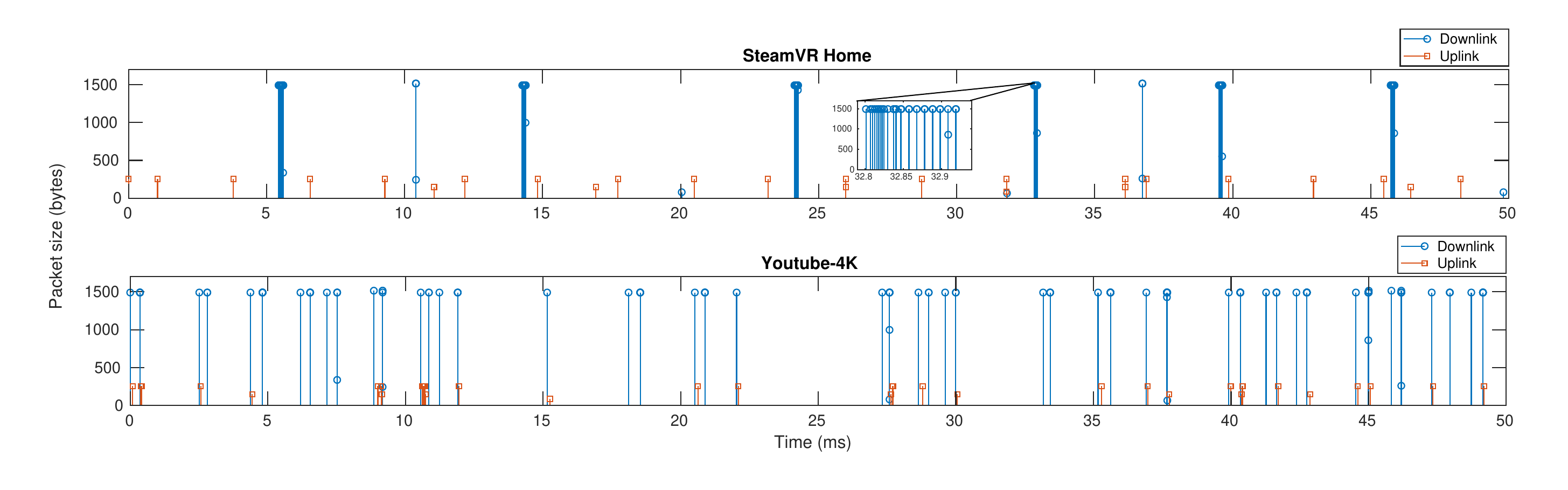}  
              \caption{Traffic traces for VR (SteamVR Home) and Non-VR (Youtube-4K) over 50ms. In order to highlight the presence of multiple packets in each batch, zoomed-in view of one VR DL batch has been illustrated.}
          \label{fig:figure2}
        \end{figure*}       
    \begin{itemize}
        \item \textbf{VR Traffic Traces Description:} for better understanding of the Interactive VR traffic data traces, a plot of SteamVR Home (120fps, 40Mbps) for a 50ms time interval, is presented in the upper plot of Fig. \ref{fig:figure2}. Most UL traces are around 254 bytes, which are related to user tracking data. The other smaller UL values of traces can be related to the statistics of HMD. On the DL side, fragments with traces 1490 bytes are common, representing the larger data packets associated with video batches. In the plot, six video batches are depicted, each containing fragments of 1490 bytes. One batch is zoomed in for clarity to highlight packets with this similar size. Besides, within each batch, there is a smaller-size packet, representing the residual packet resulting from frame fragmentation, which is smaller than a full fragment.
        \item \textbf{Non-VR Traffic Traces Description:} A plot of the Non-VR data traces in YouTube (as an example) for a 50ms time interval, is depicted in the lower plot of Fig. \ref{fig:figure2}. YouTube primarily streams pre-recorded video content, which is relatively consistent in terms of data size, driven by video streaming requirements and it is not interactive.  As shown in the figure, UL packets are around 80 bytes, representing client video requests, while DL packets are around 1290 bytes, related to video streaming that is sent to the client. 
    \end{itemize}

    \item \textbf{Feature Engineering:} in this section, feature extraction and feature selection are employed to create features that are useful for the subsequent classification process, as discussed below:
    \begin{itemize}
        \item \textbf{Feature Extraction:} the dataset contains Wireshark traces with information about the exchanged data packets. In the initial step of our analysis, we define some short periods of time. Each set of packets within a period is called a `sample'. For each sample, first, we separate the DL and UL data, and then compute Number of Packets, Total Bytes and statistics such as Minimum, Maximum, Mean, and Standard Deviation for Packet Size and Packet Inter-arrival Time, for both DL and UL, separately. In general, VR services follow some common and repetitive patterns in the DL and UL directions. Incorporating features that consider this distinct relationship or correlation between DL and UL can contribute to a more robust analysis and accurate identification of VR and Non-VR traffic. Therefore, we also compute three additional features considering this correlation. The first one is the Ratio of Number of Packets (derived from the division of Number of Packets in DL by Number of Packets in UL). The Ratio of Total Bytes (calculated from the division of Total Bytes in DL by Total Bytes in UL) is considered as a second feature in each sample for further analysis. Furthermore, in each sample we incorporate the cross-correlation among total bytes in a group of DL and UL traffic as a third feature. To compute the cross-correlation for each sample, we divide the duration of each sample into a certain number of sub-samples, resulting in $\tau$, according to the next expression (\ref{eq:equation1}):
        \begin{equation} \label{eq:equation1}
            \tau = \frac{\omega}{N}\ 
        \end{equation}
        where $\omega$ designates the duration of the sample considered in millisecond (ms) and $N$ denoted to number of sub-samples. Fig. \ref{fig:figure3} illustrates the partitioning of a sample within a duration $\omega$ into a designated set of $N$ sub-samples.

        The total number of bytes per sub-sample ($\text{Sub-sample}_n$) can be obtained by the next expression (\ref{eq:equation2}):
        \begin{equation}
            \rho_n = \sum_{p=1}^{P}S_p 
            \label{eq:equation2}
        \end{equation}
        where $\rho_n$ denotes total sum of packet sizes $S_p$ considered in the duration of $\tau$ of $\text{Sub-sample}_n$, where $n = \{1, 2, …, N\}$ and $p = \{1, 2, …, P\}$ being P the total number of packets in a duration of $\tau$ of $\text{Sub-sample}_n$. The expression (\ref{eq:equation2}) can be used for obtaining  $\rho_n$ for both DL and UL. These sub-samples can be gathered in two vectors, designated as $D = \{\rho_n\}$ for DL and $U = \{\rho_n\}$ for UL. Therefore, the Pearson correlation \cite{brownlee2018statistical} is calculated according to (\ref{eq:equation3}).

        
        
        The $C_{D,U}$ is the cross correlation between $D$ and $U$ considered as a feature 
        for each sample. We obtain the correlation values for all samples in each dataset.      
        \begin{strip}
                \begin{equation}
                    C_{D,U} = \frac{N(\sum_{j=1}^{N}D_{j}U_{j}) - (\sum_{j=1}^{N}D_{j})(\sum_{j=1}^{N}U_{j})}{\sqrt{[N\sum_{j=1}^{N}D_{j}^2-(\sum_{j=1}^{N}D_{j})^2][N\sum_{j=1}^{N}U_{j}^2-(\sum_{j=1}^{N}U_{j})^2]}}\ 
                    \label{eq:equation3}
                \end{equation}
        \end{strip} 
        Table \ref{tab:table2} shows the considered symbols for the 23 features extracted for each sample and a brief description of each. It is worth noting that, for the sake of simplicity in the paper, we have assigned a symbol to each feature, which can be found in Table \ref{tab:table2} for reference.
                
        In addition to extracted features, a binary label is also included in the dataset to identify VR from Non-VR (i.e., assigning 1 to VR and 0 to Non-VR).

        \begin{table}[t]
          \caption{The Description of Symbols for Features Extracted. Ten similar features are considered for DL and UL separately, distinguished by `DL' or `UL' at the end of their respective symbols.}
          \centering
          \scriptsize 
          \setlength{\tabcolsep}{4pt} 
          
          \resizebox{\columnwidth}{!}{
              \begin{tabular}{| c | c |p{1cm}|}      
                \hline
                 Feature Symbol & Description (per $\omega$) \\
            
                \hline
                 NoPDL & \multirow{2}{*}{Number of packets}\\
                 NoPUL &  \\
                \hline
                  TBDL  & \multirow{2}{*}{Total Bytes}  \\
                  TBUL & \\
                \hline
                  MinPSDL & \multirow{2}{*}{Min Packet Size} \\
                  MinPSUL & \\
                \hline
                 MaxPSDL & \multirow{2}{*}{Max Packet Size} \\ 
                 MaxPSUL &  \\  
                \hline
                 MeanPSDL & \multirow{2}{*}{Mean Packet Size} \\ 
                 MeanPSUL &  \\      
                \hline
                 StdPSDL & \multirow{2}{*}{Standard Deviation Packet Size} \\ 
                 StdPSUL &  \\
                 \hline
                 MinPIATDL & \multirow{2}{*}{Min Packet inter-arrival time} \\ 
                 MinPIATUL &  \\ 
                 \hline
                 MaxPIATDL & \multirow{2}{*}{Max Packet inter-arrival time} \\
                 MaxPIATUL & \\ 
                 \hline
                  MeanPIATDL & \multirow{2}{*}{Mean Packet inter-arrival time} \\
                  MeanPIATUL &  \\ 
                 \hline
                 StdPIATDL & \multirow{2}{*}{Standard Deviation Packet inter-arrival time} \\ 
                 StdPIATUL &  \\ 
                \hline
                 RoNoP & Ratio of Number of Packets (DL/UL) \\ 

                 \hline
                 RoTB & Ratio of Total Bytes (DL/UL)\\
                 
                \hline
                  CC & Cross Correlation (i.e. $C_{D,U}$)\\ 
                 
                \hline
              \end{tabular}
          }
          \label{tab:table2}
      \end{table}

      \begin{figure}[t]
          \centering
              \includegraphics[width=1\linewidth]{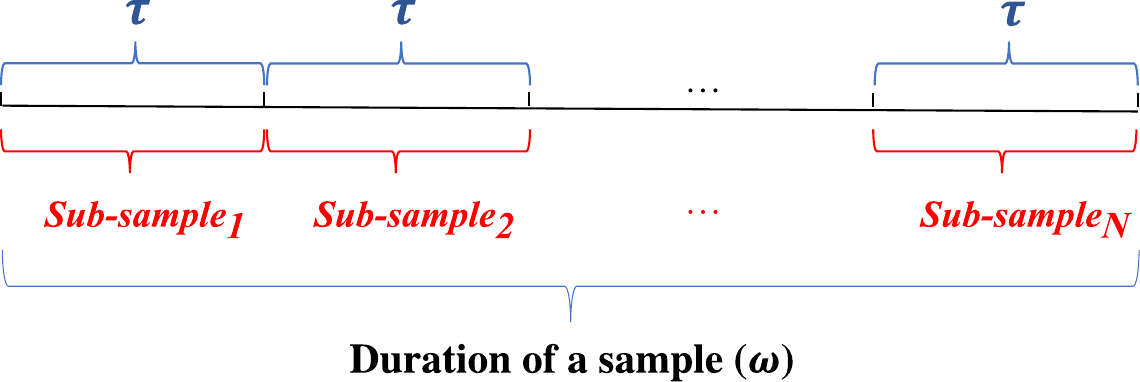}  
              \caption{The division of $\omega$ into a specified $N$ number of sub-samples.}
          \label{fig:figure3}
    \end{figure} 
        
        \item \textbf{Feature Selection:} there are several feature selection techniques in supervised learning, like  ``filter" that selects feature subsets based on their relationship with the target variable by statistical methods or feature importance methods, ``wrapper" that searches for effective feature subsets \cite{azab2022network, kuhn2013applied}, ``intrinsic" that selects by algorithms like Decision Trees to perform automatic feature selection during training \cite{kuhn2013applied} or some novel ``hybrid models", combining them \cite{cao2020improved}. We use filter selection using a feature importance method, called ``permutation importance" that is both model-agnostic and easily comprehensible. Permutation importance stands out as it possesses both of these qualities. It assesses feature importance by quantifying the impact on model error (e.g., Mean Absolute Error (MAE), r-squared, accuracy) when the values of a single feature are permuted, providing an intuitive and straightforward measure \cite{pedregosa2011scikit}. Based on permutation importance feature selection technique, the importance of each feature has been obtained in each specific dataset with a single sample duration, specific correlation sub-sample and for each ML classification algorithm. For most of these algorithms, it is clear that using all the features is crucial to achieve high performance. Nevertheless, there are instances where certain features have been disregarded by the algorithms, such as the Decision Trees (DTs) classifier in all varied settings.

    \end{itemize}
\end{enumerate}


\subsection{Traffic Classification Methods}
\label{subsecclassmethod}
Traffic classification methods are essential for managing and optimizing network performance by identifying the types of traffic flowing through the network. In this section, we first present a background of existing traffic classification methods, highlighting the comparison among them. Then, we describe how we apply ML to classify traffic in our dataset.

\subsubsection{Traffic Classification Background}

Different kinds of traffic classification methods can be found in the literature. Traditional methods are based on port-based or payload-based (DPI). On the other hand, other traffic classification methods are based on AI/ML. Table \ref{tab:tablecompare} compares these three methods according to several aspects including complexity, computational time, handling encrypted traffic, and adaptability. The traditional methods require manual tuning, whereas the ML-based approach works seamlessly after training. Moreover, port-based classification is simple to implement, requires minimal computing resources, and offers high-speed classification. However, it involves accessing the packet's header and examining the utilized port number, providing a simple method with low computational demands. Payload-based (DPI) achieves higher classification accuracy by analyzing the content of packets within the monitored flow. Furthermore, DPI methods are ineffective in classifying encrypted network traffic because encryption obscures the payload, making it unreadable without the decryption keys, which are not accessible due to privacy and security reasons. This issue reduces their utility as many applications utilize encryption. Moreover, DPI's processing speed is slower, particularly when handling aggregated packets. In contrast, ML-based, specifically Supervised Learning (SL)-based classification methods, provide high accuracy by learning from data patterns and moderate computational time, handle encrypted traffic well by analyzing metadata and traffic patterns, and offer fine-grained detection without needing to access the content of network traffic, thus making them highly adaptable to new scenarios without manual intervention. ML-based techniques require moderate complexity for feature extraction and model training \cite{azab2022network}.

\begin{table}[t]
  \caption{Comparison of Traffic Classification Methods.}
  \centering
  \scriptsize 
  \setlength{\tabcolsep}{4pt} 
  
  \resizebox{\columnwidth}{!}{
      \begin{tabular}{| c | c | c | c |p{1cm}|}   
        \hline
         Aspect & Port-based & Payload-based & ML-based \\
        \hline
         Complexity & Low & High & Moderate \\
        \hline
         Computational & \multirow{2}{*}{Low} & \multirow{2}{*}{High} & \multirow{2}{*}{Moderate} \\
         Time &  &  &  \\
        \hline
         Handling & \multirow{2}{*}{Yes} & \multirow{2}{*}{No} & \multirow{2}{*}{Yes} \\
         Encrypted Traffic &  &  &  \\ 
        \hline
         Adaptability & Low & Low & High \\      
        \hline
      \end{tabular}
  }
  
  \label{tab:tablecompare}
\end{table}

\subsubsection{ML-based Traffic Classification}
For traffic classification, we aim to identify simple classifiers that can be later easily implemented in the AP. As such, six common and distinct ML classifiers, namely Logistic Regression (LR), Support Vector Machines (SVM), k-Nearest Neighbors (kNN), Decision Trees (DT), Random Forest (RF), and Naive Bayes (NB) \cite{brownlee2016machine} are selected. A brief background on these classifiers and their common hyperparameters are provided below.

\begin{enumerate}
    \item \textbf{Background:} a concise overview of each classifier used in our study is provided in the following: 
    \begin{itemize}
            \item \textbf{LR:} a linear classification algorithm used for binary or multi-class classification that models the probability of an instance belonging to a particular class using a logistic function. It is simple, interpretable, and works well when the decision boundary is linear \cite{brownlee2016machine, charu2020data}.
            \item \textbf{SVM:} a versatile algorithm for classification that finds a hyperplane for best separating data points of different classes in classification. It is effective in high-dimensional spaces and can handle non-linear data using kernel functions \cite{azab2022network, brownlee2016machine, charu2020data}. 
            \item \textbf{kNN:} a simple instance-based learning algorithm that classifies data points based on the majority class among their k nearest neighbors. It is intuitive and works well when data is locally clustered \cite{azab2022network, brownlee2016machine, charu2020data}.
            \item \textbf{DT:} a tree-like structure where each node represents a decision based on a feature. They are interpretable and can capture non-linear relationships in the data. However, they are prone to overfitting \cite{azab2022network, brownlee2016machine, charu2020data}.
            \item \textbf{RF:} an ensemble method that consists of multiple decision trees that improves upon the weaknesses of individual decision trees by aggregating their predictions. RF is robust, handles high-dimensional data well, and reduces overfitting \cite{azab2022network, brownlee2016machine, charu2020data}. 
            \item \textbf{NB:} a probabilistic classification algorithm based on Bayes’ theorem that assumes independence between features (a ``naïve” assumption) and calculates the probability of an instance belonging to a class. NB is simple and efficient \cite{azab2022network, brownlee2016machine, charu2020data}.
    \end{itemize}
    \item \textbf{Hyperparameter Description:} in our case, we apply these supervised learning classifiers for binary classification, assigning labels as either ``VR" or ``Non-VR" traffic. All classification models are constructed using the scikit-learn \cite{pedregosa2011scikit} machine learning library. During the classification process, we perform hyperparameter tuning for ML models using GridSearchCV \cite{pedregosa2011scikit, brownlee2016machine} for each classifier. GridSearch is employed to identify the optimal hyperparameters for each classifier. Below, we provide brief descriptions of the common hyperparameters for each classifier:
    \begin{itemize}
            \item \textbf{LR and SVM:} in both Logistic Regression (LR) and Support Vector Machines (SVM), a key hyperparameter is ``C", representing the regularization strength. ``C" controls the degree of regularization applied to the model and inversely influences its level of regularization. However, there are specific hyperparameters unique to each algorithm. In LR, we have the ``solver" hyperparameter, which dictates the optimization algorithm used during model training. Meanwhile, SVM introduces the ``kernel" hyperparameter, determining the kernel function applied to the input data to facilitate the discovery of non-linear decision boundaries. This transformation is vital for SVMs to capture complex relationships within the data.
            \item \textbf{kNN:} in k-Nearest Neighbors (kNN), two essential hyperparameters are ``n\_neighbors" and ``weights." ``\text{n\_neighbors}” determines the number of nearest neighbors that the algorithm considers when making a prediction for a new data point and ``weights” defines the weight assigned to each neighbor when making a prediction. The ``weights" hyperparameter in kNN can take on one of two common values: ``uniform," which assigns equal weight to all neighbors in the prediction, or ``distance", which assigns weights inversely proportional to each neighbor's distance from the data point being predicted.
            \item \textbf{DT and RF:} for DT and RF, one of the hyperparameters is ``max\_depth” that signifies the maximum depth of a tree. When left unspecified, the tree expands until each leaf node contains only one value. Therefore, by reducing this parameter, we can prevent the tree from learning all training samples, thus mitigating the risk of overfitting. In addition, among the other hyperparameters in DT and RF, such as min\_samples\_leaf, max\_leaf\_nodes, and min\_impurity\_decrease, which allow for the development of asymmetric trees and impose constraints at the leaf or node level, we consider the impact of ``min\_samples\_split”. In the case of only Random Forest (RF), the hyperparameter ``n\_estimator" is also taken into account to restrict the number of decision trees in the ensemble. 
            \item \textbf{NB:} the ``var\_smoothing" hyperparameter in Naive Bayes (NB) algorithms, controls the amount of smoothing applied to the variance of numerical features. Smoothing helps prevent issues when calculating probabilities, especially for features that have zero variance in the training data. 
    \end{itemize}
\end{enumerate}


\section{Results} \label{secobtresults}

This section first describes the obtained results for the extracted features and the corresponding files generated based on them, which serve as input for the classification process. Second, it presents the results of traffic classification.


\subsection{Feature Extraction}

We define $\omega$ with values of 1000ms (1 second), 500ms, 200ms, 100ms, and 50ms. From initial packet traces, for different values of sample duration ($\omega$), we extract features, sample by sample, and store in a single dataset, resulting in a total of five datasets. We consider only samples that contain a set of non-zero packets. Table \ref{tab:table3} illustrates the count of obtained VR samples for VR, Non-VR samples and  total number of samples when the sample duration ($\omega$) takes different values. In the training phase of the classification process, it is essential to strive for a balanced distribution of data across labels. We can observe inconsistencies between the number of VR and Non-VR samples as we consider smaller values of $\omega$. VR applications demand rapid updates and real-time interactions, leading to more frequent transmission of smaller data chunks and therefore, even for low $\omega$, the number of samples with zero packets is low. On the other hand, Non-VR data show more samples with zero packet traces than VR due to their own nature. To address this, we removed the last samples of VR to balance the data distribution across VR and Non-VR labels. This removal process was applied to all dataset except dataset with $\omega=1 sec$, and in the case of $\omega=500ms$, fewer number of samples were removed. As we approach about $\omega=50ms$, a larger number of samples needs to be removed to maintain balance.
\begin{table}
  \caption{The count of obtained VR samples, Non-VR samples and total number of samples for different values of sample duration ($\omega$). Note that shorter sample duration leads to higher number of samples.}
  \centering
  \scriptsize 
  \setlength{\tabcolsep}{4pt} 
  
  \resizebox{\columnwidth}{!}{
      \begin{tabular}{| c | c | c | c |p{1cm}|}   \hline
         \multirow{2}{*}{$\omega$} & VR Traffic & Non-VR Traffic & Total \\
          & samples & samples & samples \\
        \hline
         1sec & 790 & 841 & 1631 \\          
        \hline
         500ms  & 1359 & 1291 & 2650 \\
        \hline
         200ms & 2421 & 2435 & 4856\\ 
        \hline
         100ms & 4176 & 4237 & 8413\\ 
          
        \hline
         50ms & 7694 & 7790 & 15484\\      
        \hline
      \end{tabular}
  }
  \label{tab:table3}
\end{table}

In order to calculate the CC feature, we define the values of 5, 10, and 20 for $N$. Therefore, for each specific values of $\omega$ and $N$ (referred to as a setting in the paper), we store the obtained features in separate dataset, resulting in a total of 15 datasets including both VR and Non-VR samples for each of dataset. Table \ref{tab:table4} displays the average CC obtained from VR samples and Non-VR samples for different values of $\omega$ and $N$ (i.e. settings).

\begin{table}
  \caption{Average CC obtained from VR samples and Non-VR samples for different values of sample duration ($\omega$) and number of sub-samples ($N$) (i.e. different settings).}
  \centering
  \scriptsize 
  \setlength{\tabcolsep}{4pt} 
  
  \resizebox{\columnwidth}{!}{
      \begin{tabular}{| c | c | c | c | c | p{1cm}|}       \hline
          \multirow{2}{*}{$\omega$} & Type of samples  & \multicolumn{3}{|c|}{Average of CC Values per $n$}\\   
         \cline{3-5}
          & (No. of samples) & 5 & 10 & 20\\
         \hline
         \multirow{2}{*}{1sec} & VR (790) & 0.96889 & 0.95294 & 0.92365\\
         \cline{2-5}
          & Non-VR (841) & 0.84653 & 0.80525 & 0.71961\\
         \hline
         \multirow{2}{*}{500ms} & VR (1359) & 0.96321 & 0.93112 & 0.84515\\
         \cline{2-5}
          & Non-VR (1291) & 0.83223 & 0.75383 & 0.66058\\
         \hline  
         \multirow{2}{*}{200ms} & VR (2421) & 0.93459 & 0.86528 & 0.76849\\
         \cline{2-5}
          & Non-VR (2435) & 0.78340 & 0.63742 & 0.48915\\
         \hline
         \multirow{2}{*}{100ms} & VR (4176) & 0.85400 & 0.75078 & 0.55645\\
         \cline{2-5}
          & Non-VR (4237) & 0.66973 & 0.50797 & 0.40811\\
         \hline
         \multirow{2}{*}{50ms} & VR (7694) & 0.74553 & 0.55422 & 0.33449\\
         \cline{2-5}
          & Non-VR (7790) & 0.53264 & 0.42503 & 0.33813\\
         \hline
      \end{tabular}
  }
  \label{tab:table4}
\end{table}

In Table \ref{tab:table4}, it is observed that the average correlation of VR traffic is consistently higher than that of Non-VR traffic across all settings with the exception of the $\omega=50ms$ and $N=20$, where they exhibit nearly identical levels of correlation. As we decrease the $\omega$, the correlation values tend to decrease. Moreover, as the number of sub-samples used to calculate the correlation increases, the correlation values decrease. As Table \ref{tab:table4} reveals, within $\omega=50ms$ and $N=20$, a narrowing gap exists between the correlations for VR and Non-VR traffic.


\subsection{Traffic Classification Results}

In this process, we evaluate all six classifiers for each of the 15 different input datasets, one per setting. Before the classification process, the dataset is split into 70\% for training and 30\% for validation. Within the classification phase, hyperparameter tuning and feature selection are conducted as follows:
\begin{enumerate}
    \item \textbf{Hyperparameter Tuning:} The considered hyperparameters by using GridSearchCV framework, which incorporates a robust cross-validation approach with a fixed count of three, are listed as below:
    \begin{itemize}
        \item ``LR\_parameters": [{``solver": [`liblinear', `saga'], `C': [0.1, 1]}]
        \item ``SVM\_parameters": [{`kernel': [`rbf',`sigmoid'],  `C': [0.1, 1]}]
        \item ``kNN\_parameters": [{`n\_neighbors': [5,10], `weights': [`uniform', `distance']}]
        \item ``DT\_parameters": [{`min\_samples\_split': [5,8], `max\_depth': [5,10]}]
        \item ``RF\_parameters": [{`n\_estimators': [5,20,50], `min\_samples\_split': [5,8], `max\_depth': [5,10]}]
        \item ``NB\_parameters": [{`var\_smoothing': np.logspace(0,-9, num=100)}]
    \end{itemize}
    \item \textbf{Permutation Importance Feature Selection:} using this technique, we select the most crucial features to influence the traffic classification process according to the specific ML classification algorithm under consideration. Features with an importance rate of zero are excluded from participation in the classification process.
\end{enumerate}

By incorporating these two processes, we perform classification method. Four commonly used metrics, namely Accuracy, Precision, Recall, and F1-Score, are employed to evaluate the effectiveness of the built classifiers in a supervised classification problem \cite{azab2022network, christen2012evaluation}. Accuracy indicates the overall performance of the model by dividing the correctly classified samples by the total number of samples in a dataset. Precision assesses the performance of the classifier in classifying samples into a specific class. Recall presents the performance of the classifier regarding the number of samples correctly identified from the entire dataset. F1-Score represents the harmonic mean of precision and recall \cite{azab2022network}. In light of this, we present a comparison of the accuracy results of classifiers during the validation phase referred to as the validation score in the paper, for each specific $\omega$ and $N$ (i.e. each of 15 settings) in Table~\ref{tab:table5}. In the context of correlation analysis, as depicted in Table~\ref{tab:table4}, a subset of 5 sub-samples exhibited superior correlation outcomes than 10 or 20 sub-samples, while as indicated in Table \ref{tab:table5}, the best validation score pertains to sample duration of 500ms (i.e. $\omega = 500ms$) and the utilization of 20 sub-samples (i.e. $N=20$) for the correlation analysis across five distinct classifiers, namely Support Vector Machines (SVM), k-Nearest Neighbors (kNN), Decision Trees (DT), Random Forest (RF), and Naive Bayes (NB). The most favorable outcome is attributed to the RF classifier, yielding a validation score of 0.99245. However, when evaluating Logistic Regression (LR), a slightly better performance is observed with sample duration of 500ms and the 20 number of sub-samples dataset, although the differences in accuracy between these values are minimal. It is important to note that while the average correlation values (as presented in Table \ref{tab:table4}) are higher for the 5 and 10 considered sub-samples compared to the 20 number of sub-sample dataset, the results indicate that classifiers exhibit a better ability to distinguish between VR and Non-VR in the 20 number of sub-sample dataset.

\begin{table}
  \caption{Validation scores of six classifiers for all considered setting (i.e. different values of $\omega$ and $N$). The highest accuracies are highlighted in bold.}
  \centering
  \setlength{\tabcolsep}{4pt} 
  
  \resizebox{\columnwidth}{!}{
      \begin{tabular}{| c | c | c | c |c | c | c | c |p{1cm}|}     
         \hline
         $N$ & $\omega$ & LR & SVM & kNN & DT & RF & NB \\
         \hline
        \multirow{5}{*}{20} & 1sec & 0.9388 & 0.9326 & 0.9306 & 0.9265 & 0.9347 & 0.8939 \\          
        \cline{2-8}
         & 500ms & 0.9597 & \textbf{0.9811} & \textbf{0.9624} & \textbf{0.9874} & \textbf{0.9924} & \textbf{0.9459} \\
        \cline{2-8}
         & 200ms & 0.9547 & 0.9547 & 0.9588 & 0.9547 & 0.9554 & 0.8881 \\ 
        \cline{2-8}
         & 100ms & 0.9616 & 0.9596 & 0.9612 & 0.9592 & 0.9592 &  0.8918 \\ 
        \cline{2-8}
         & 50ms & \textbf{0.9621} & 0.9578 & 0.9587 & 0.9597 & 0.9632 & 0.9053 \\
        \hline
        \multirow{5}{*}{10} & 1sec & 0.9224 & 0.9326 & 0.9265 & 0.9286 & 0.9306 & 0.8918 \\ 
        \cline{2-8}
         & 500ms & 0.9572 & 0.9597 & 0.9610 & 0.9572 & 0.9648 & 0.9270 \\
        \cline{2-8}
         & 200ms & 0.9526 & 0.9561 & 0.9554 & 0.9513 & 0.9513 & 0.8964 \\ 
        \cline{2-8}
         & 100ms & 0.9544 & 0.9564 & 0.9552 & 0.9576 & 0.9584 & 0.8930 \\ 
        \cline{2-8}
         & 50ms & 0.9537 & 0.9529 & 0.9518 & 0.9507 & 0.9546 & 0.9029 \\
        \hline
        \multirow{5}{*}{5} & 1sec & 0.9286 & 0.9245 & 0.9265 & 0.9245 & 0.9265 & 0.9122 \\ 
        \cline{2-8}
         & 500ms & 0.9610 & 0.9535 & 0.9459 & 0.9522 & 0.9535 & 0.9157 \\
        \cline{2-8}
         & 200ms & 0.9554 & 0.9574 & 0.9581 & 09574 & 0.9581 & 0.8888 \\ 
        \cline{2-8}
         & 100ms & 0.9481 & 0.9485 & 0.9497 & 0.9473 & 0.9509 & 0.8902 \\ 
        \cline{2-8}
         & 50ms & 0.9546 & 0.9565 & 0.9574 & 0.9582 & 0.9593 & 0.8969 \\
        \hline
      \end{tabular}
  }
  \label{tab:table5}
\end{table}

In Table \ref{tab:table6}, we provide the validation report for each classifier under the best-performing setting, which corresponds to sample duration of 500ms and number of sub-samples of 20 (i.e. $\omega=500$ms and $N=20$). In the support column, indicating the actual occurrences of each class in the specified dataset, it is evident that 795 samples are included in the validation process to assess the training data, comprising 392 for Non-VR and 403 for VR. Regarding the Precision, Recall, and F1-Score metrics for the top three classifiers (kNN, DT, and RF), all values stand at 0.98 and 0.99, with an increase in the precision of VR in the top-performing classifier, RF, which is higher than 0.995 and rounds to 1.00.

\begin{table}
  \caption{Validation report for the setting of $\omega=500ms$ and $N=20$.}
  \centering
  \scriptsize 
  \setlength{\tabcolsep}{4pt} 
  
  \resizebox{\columnwidth}{!}{
      \begin{tabular}{| c | c | c | c | c | c |p{1cm}|}     
         \hline
         \multirow{2}{*}{Classifier} & Type of & \multirow{2}{*}{Precision} & \multirow{2}{*}{Recall} & \multirow{2}{*}{F1-Score} & \multirow{2}{*}{Support} \\
         & Traffic &  &  &  &  \\
        \hline
        \multirow{2}{*}{LR} & Non-VR & 0.94 & 0.98 & 0.96 & 392 \\          
        \cline{2-6}
         & VR & 0.98 & 0.94 & 0.96 & 403 \\
        \hline
         \multirow{2}{*}{SVM} & Non-VR & 0.97 & 0.99 & 0.98 & 392 \\          
        \cline{2-6}
         & VR & 0.99 & 0.98 & 0.98 & 403 \\
        \hline
        \multirow{2}{*}{kNN} & Non-VR & 0.98 & 0.98 & 0.98 & 392 \\          
        \cline{2-6}
         & VR & 0.98 & 0.99 & 0.98 & 403 \\
        \hline
        \multirow{2}{*}{DT} & Non-VR & 0.99 & 0.98 & 0.99 & 392 \\          
        \cline{2-6}
         & VR & 0.99 & 0.99 & 0.99 & 403 \\
        \hline
        \multirow{2}{*}{RF} & Non-VR & 0.99 & 0.99 & 0.99 & 392 \\          
        \cline{2-6}
         & VR & 1.00 & 0.99 & 0.99 & 403 \\
        \hline
        \multirow{2}{*}{NB} & Non-VR & 0.99 & 0.90 & 0.94 & 392 \\          
        \cline{2-6}
         & VR & 0.91 & 0.99 & 0.95 & 403 \\
        \hline
      \end{tabular}
  }
  \label{tab:table6}
\end{table}

Table \ref{tab:table7} displays the confusion matrices depicting true labels and predicted labels for all classifiers in the best-performing setting. In the case of the top-performing classifier, RF, it correctly labels 390 out of 392 Non-VR samples, with only 2 misclassified as VR. For the 403 VR samples, it accurately predicts 399 as true VR and misclassifies 4 of them. On the other hand, in the case of NB, which exhibits lower accuracy metrics compared to the others, we observe that this performance is primarily influenced by the misclassification of 39 Non-VR traffic instances as VR.

\begin{table}
  \caption{Confusion matrices of validation data for the setting of $\omega=500ms$ and $N=20$.}
  \centering
  \scriptsize 
  \setlength{\tabcolsep}{4pt} 
  
  \resizebox{\columnwidth}{!}{
      \begin{tabular}{| c | c | c | c | c | p{1cm}|}     
         \hline
         \multirow{2}{*}{} & \multirow{2}{*}{Classifier} & \multirow{2}{*}{Type of Traffic} & \multicolumn{2}{|c|}{Predicted Label}  \\
         \cline{4-5}
         & &  & Non-VR & VR \\
        \hline
        \multirow{12}{*}{\rotatebox[origin=c]{90}{True Label}} & \multirow{2}{*}{LR} & Non-VR & 384 & 8 \\ 
        \cline{3-5}
         &  & VR & 24 & 376 \\ 
        \cline{2-5}
         & \multirow{2}{*}{SVM} & Non-VR & 387 & 5 \\ 
        \cline{3-5}
         &  & VR & 10 & 393 \\ 
        \cline{2-5}
        & \multirow{2}{*}{kNN} & Non-VR & 384 & 8 \\ 
        \cline{3-5}
         &  & VR & 6 & 397 \\ 
        \cline{2-5}
        & \multirow{2}{*}{DT} & Non-VR & 386 & 6 \\ 
        \cline{3-5}
         &  & VR & 4 & 399 \\ 
        \cline{2-5}
        & \multirow{2}{*}{RF} & Non-VR & 390 & 2 \\ 
        \cline{3-5}
         &  & VR & 4 & 399 \\ 
        \cline{2-5}
        & \multirow{2}{*}{NB} & Non-VR & 353 & 39 \\ 
        \cline{3-5}
         &  & VR & 4 & 399 \\ 
        \hline
        
      \end{tabular}
  }
  \label{tab:table7}
\end{table}

Upon reviewing all the results, we assessed the importance of the features for the top three classifiers in the best-performing setting. This was done to gauge the significance of each feature in the process of permutation importance feature selection. Table \ref{tab:table8} provides a breakdown of the importance of each feature for each classifier in sequential order. It is worth noting that symbols representing the features are used in this table to simplify and streamline the presentation. In the best-performing setting for these three classifiers, we can observe the order of feature importance that yielded the best results. In the case of kNN, all features are utilized in the classification process, whereas the DT classifier employs only 12 features. Moreover, in DT, the first feature ranked in importance is RoNoP, with a high value (i.e., 0.7751) which can present the significance of the correlation between DL and UL to distinguish VR from Non-VR. In RF, all features except for ``MinPIATDL" are utilized in the classification process.  

\begin{table*}[t]
  \caption{Importance of the features in three top-performing classifiers for the best-performing setting (i.e. $\omega=500ms$ and $N=20$).}
  \centering
  \small
  \setlength{\tabcolsep}{4pt} 
  
      \begin{tabular}{| c | c | c | c | c | c | c | c | c | p{1cm}|}       \hline
         \multirow{3}{*}{No.} & Features & Importance & \multirow{3}{*}{No.} & Features & Importance & \multirow{3}{*}{No.} & Features & Importance \\
          & ranked & values &  & ranked & values &  & ranked & values \\
          & importance & for kNN &  & importance & for DT &  & importance & for RF \\
         \hline
         0 & NoPDL & 0.0930 & 0 & RoNoP & 0.7751 & 0 & MaxPSUL & 0.2155 \\
         \hline
         1 & CC & 0.0894 & 1 & MaxPSUL & 0.1484 & 1 & MaxPIATDL & 0.1380 \\
         \hline
         2 & TBDL & 0.0809 & 2 & NoPUL & 0.0344 & 2 & RoNoP & 0.1199 \\
         \hline
         3 & NoPUL & 0.0445 & 3 & MinPIATUL & 0.0137 & 3 & TBDL & 0.1107 \\
         \hline
         4 & MeanPSDL & 0.0323 & 4 & TBDL & 0.0064 & 4 & MeanPSUL & 0.1000 \\
         \hline
         5 & TBUL & 0.0216 & 5 & MaxPIATUL & 0.0051 & 5 & MeanPIATDL & 0.0796 \\
         \hline
         6 & MeanPSUL & 0.0163 & 6 & StdPIATUL & 0.0048 & 6 & TBUL & 0.0473 \\
         \hline
         7 & MaxPSDL & 0.0079 & 7 & MinPSUL & 0.0038 & 7 & StdPIATDL & 0.0374 \\
         \hline
         8 & RoNoP & 0.0069 & 8 & MaxPIATDL & 0.0024 & 8 & MinPSDL & 0.0348 \\
         \hline
         9 & MaxPSUL & 0.0066 & 9 & CC & 0.0021 & 9 & MeanPIATUL & 0.0313 \\
         \hline
         10 & RoTB & 0.0054 & 10 & NoPDL & 0.0020 & 10 & RoTB & 0.0299 \\
         \hline
         11 & StdPSDL & 0.0036 & 11 & TBUL & 0.0017 & 11 & NoPUL & 0.0202 \\
         \hline
         12 & MaxPIATDL & 0.0036 & 12 & MeanPSDL & 0.00003 & 12 & MaxPSDL & 0.0177 \\
         \hline
         13 & StdPSUL & 0.0030 & 13 & StdPIATDL & 0.0000 & 13 & NoPDL & 0.0040 \\
         \hline
         14 & MinPSDL & 0.0027 & 14 & MeanPIATDL & 0.0000 & 14 & MinPIATUL & 0.0037 \\
         \hline
         15 & MaxPIATUL & 0.0022 & 15 & MinPIATDL & 0.0000 & 15 & MinPSUL & 0.0033 \\
         \hline
         16 & MeanPIATDL & 0.0013 & 16 & MeanPSUL & 0.0000 & 16 & MaxPIATU & 0.0015 \\
         \hline
         17 & StdPIATDL & 0.0012 & 17 & StdPSUL & 0.0000 & 17 & StdPSDL & 0.0014 \\
         \hline
         18 & StdPIATUL & 0.0012 & 18 & StdPSDL & 0.0000 & 18 & StdPIATUL & 0.0013 \\
         \hline
         19 & MinPSUL & 0.0009 & 19 & MeanPIATU & 0.0000 & 19 & StdPSUL & 0.0012 \\
         \hline
         20 & MinPIATDL & 0.0005 & 20 & MaxPSDL & 0.0000 & 20 & MeanPSDL & 0.0011 \\
         \hline
         21 & MeanPIATUL & 0.0002 & 21 & MinPSDL & 0.0000 & 21 & CC & 0.0002 \\
         \hline
         22 & MinPIATUL & 0.0002 & 22 & RoTB & 0.0000 & 22 & MinPIATDL & 0.0000 \\
         \hline
      \end{tabular}
  \label{tab:table8}
\end{table*}
Regarding the hyperparameters selected among all those considered during GridSearchCV for the three best classifiers, we observe that setting the following values leads to stabilized accuracy metrics:
\begin{itemize}
    \item For k-Nearest Neighbors (kNN), configuring n\_neighbors as 5 and employing 'distance' as the weighting scheme.
    \item For Decision Trees (DT), setting max\_depth to 10 and min\_samples\_split to 5.
    \item For Random Forest (RF), setting max\_depth as 10, min\_samples\_split as 8, and employing 50 estimators (n\_estimators). 
\end{itemize}


\section{Assessing the Interactive VR Traffic Identification model} 
\label{secresults}


\subsection{Testing the model}

In this section, we evaluate the classification model using the packet traces from: (i) a multi-user VR session using ALVR, and (ii) a single-user session using Steam Link\footnote{https://store.steampowered.com/app/353380/Steam\_Link/}. 

\begin{figure*}[t]
  \centering
      \includegraphics[width=1\linewidth]{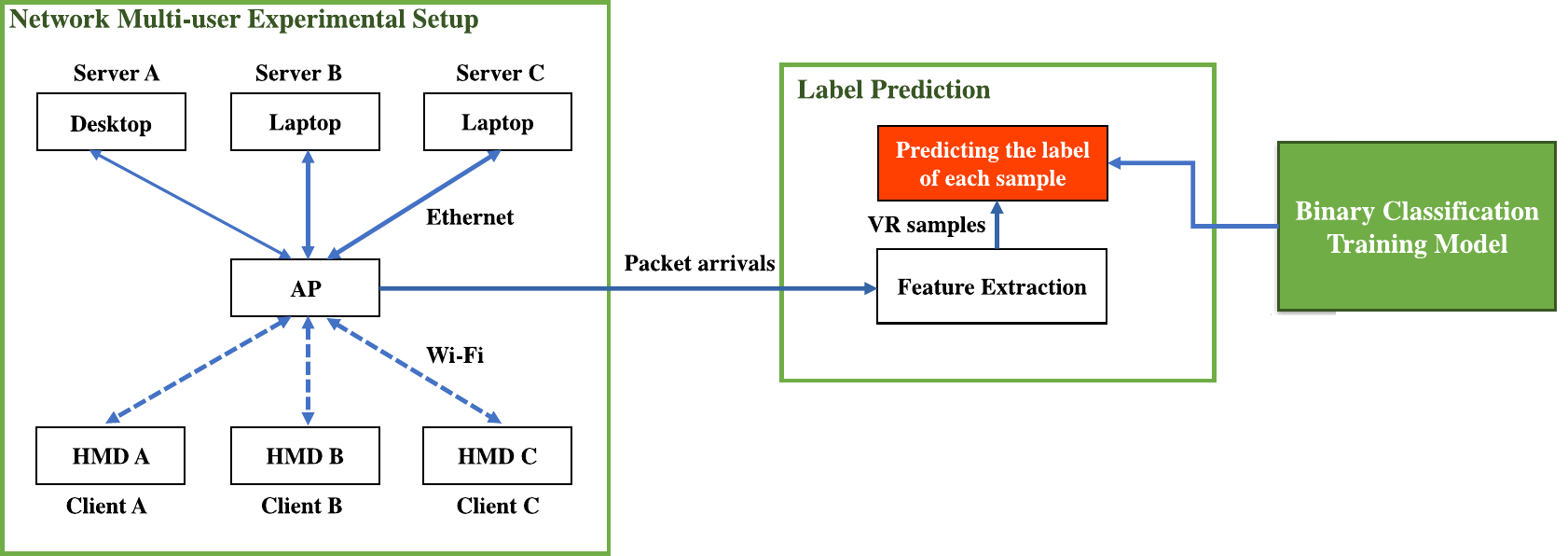}  
      \caption{Testing the Interactive VR Traffic Identification Model in a Multi-user Experimental Setup.}
  \label{fig:figure4}
\end{figure*}

First, we test the classification model with packet traces of three users playing VR games in a multi-user experimental setup~\cite{michaelides2023wi}. As shown in Fig. \ref{fig:figure4}, each computer is connected to the AP using an Ethernet cable, and each HMD is connected to the AP over Wi-Fi 6. Each HMD client is connected to a server running on a computer. During the training process, we used single-user traffic traces. The evaluation aims to ensure that the model performs effectively on multi-user data that was not part of training process. To achieve this, we evaluate the model for each user, separately. This process is illustrated in the Fig. \ref{fig:figure4}. For each user, in label prediction, first we apply the feature extraction model on packet arrivals to obtain samples in which the features are computed and then, the classification model to obtain whether the model can correctly predict its type or not. We apply two top-performing classifiers (i.e., DT and RF), for best-performing setting (i.e., $\omega = 500$ms, $N = 20$). From 128 samples of the first user (i.e., Client A) that was playing a VR game, 127 samples are correctly classified (i.e., VR), and only one is misclassified when applying the RF classifier. Therefore, the accuracy result for the test phase (referred to as test score) is: 0.9922, while with DT, 126 samples are predicted in a correct class and 2 misclassified and the test score is: 0.9844. 
For the Client B and Client C, both DT and RF classifiers achieve consistent test scores of 0.9925 for Client B and 0.9921 for Client C. Only one VR sample is misclassified by both classifiers, with 134 samples for Client B and 127 samples for Client C. 
Additionally, we assessed the computational time exclusively for Client A. In each sample, it takes approximately 0.7670 seconds for feature extraction and 0.0131 seconds for classification using a DT classifier. The computational time increases slightly to 0.0240 seconds when using a RF classifier. Thus, the total computational time (i.e. both feature extraction and classification phases) remains below 1 second for each sample. Consequently, when a user starts a game, the AP can promptly detect whether the traffic corresponds to VR or not.

Secondly, we test the classification model with packet traces from Steam Link in a single-user setup. Steam Link is the streaming solution of Valve that has been recently made available for VR. This evaluation aims to ensure that the model performs effectively with a source that was not used in the training process. Note that in the training process we only used packet traces from ALVR and Unity Render Streaming. For the traffic of a single user, we first use the feature extraction model on the packet arrivals to obtain samples with computed features. Then, we apply the two top-performing classifiers (i.e., DT and RF), for best-performing setting (i.e., $\omega = 500$ms, $N = 20$). From 123 samples of the user playing a VR game, 121 samples are correctly classified as VR, with only two misclassified by both the DT and RF classifiers. Therefore, the test phase accuracy (test score) is 0.9837. 
Furthermore, we evaluated the computational time, too. For each sample, feature extraction takes about 0.8710 seconds, and classification with a DT classifier takes 0.0137 seconds. When using an RF classifier, the classification time increases slightly to 0.0251 seconds. Therefore, the total computational time, including both feature extraction and classification, remains less than 1 second per sample.


\subsection{Wi-Fi QoS enhancement through the Prioritization of VR traffic}

\begin{figure}[t]
  \centering
      \includegraphics[width=1\linewidth]{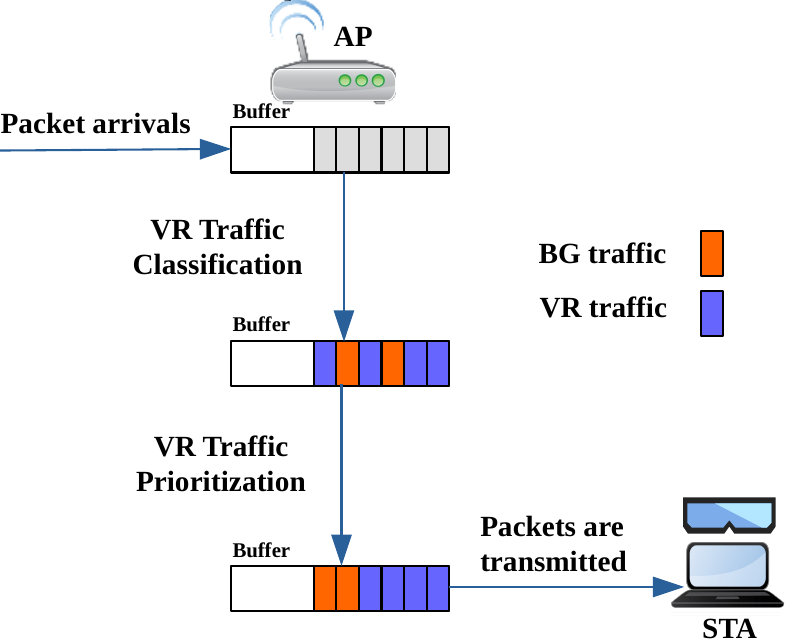}  
      \caption{System operation example: VR traffic classification and prioritization.}
  \label{fig:figure5}
\end{figure}

In this section we use a network simulator to analyze the impact that VR traffic identification can have in current Wi-Fi networks. We employ an extended version of the C++ simulator used in \cite{carrascosa2023understanding}, which allows us to simulate Wi-Fi 6 networks reproducing ALVR traces. We consider a Cloud Edge VR scenario that consists of an Access Point (AP) and two Stations (STAs). The following MCS are used: 1024-QAM 5/6 for the first STA, and 256-QAM 5/6 for the second one, both use 80 MHz channels on the 5 GHz band. The first STA is using ALVR to play a VR game at 100 Mbps and 90 fps. The second STA is receiving non-VR background (BG) ON/OFF traffic with a load ranging from 200 to 400 Mbps. The ON/OFF BG traffic alternates the ON and OFF states (exponentially distributed) for 70 ms and 30 ms in average, respectively. During the ON period, packets arrive to the AP following a Poisson distribution. To showcase the impact that traffic identification can have in Wi-Fi QoS, we allow the AP to detect whether the traffic is VR or BG, and prioritize the first one accordingly, ignoring all BG packets until all VR packets are transmitted. For best-performing setting (i.e. $\omega=500$ms and $N=20$), there are between 280 to 300 packets in each VR sample of 500 ms. After applying the classification model and determining the sample is VR, all arriving packets with the same source and destination IP will receive high priority from the AP and will be transmitted. Note that the classifier is activated once enough packets are gathered for a 500ms sample to determine the label of incoming traffic in the sample. To better illustrate the AP prioritization process, a traffic scenario for a few packets is depicted in Fig. \ref{fig:figure5}.

\begin{figure}[t]
  \centering
      \includegraphics[width=1\linewidth]{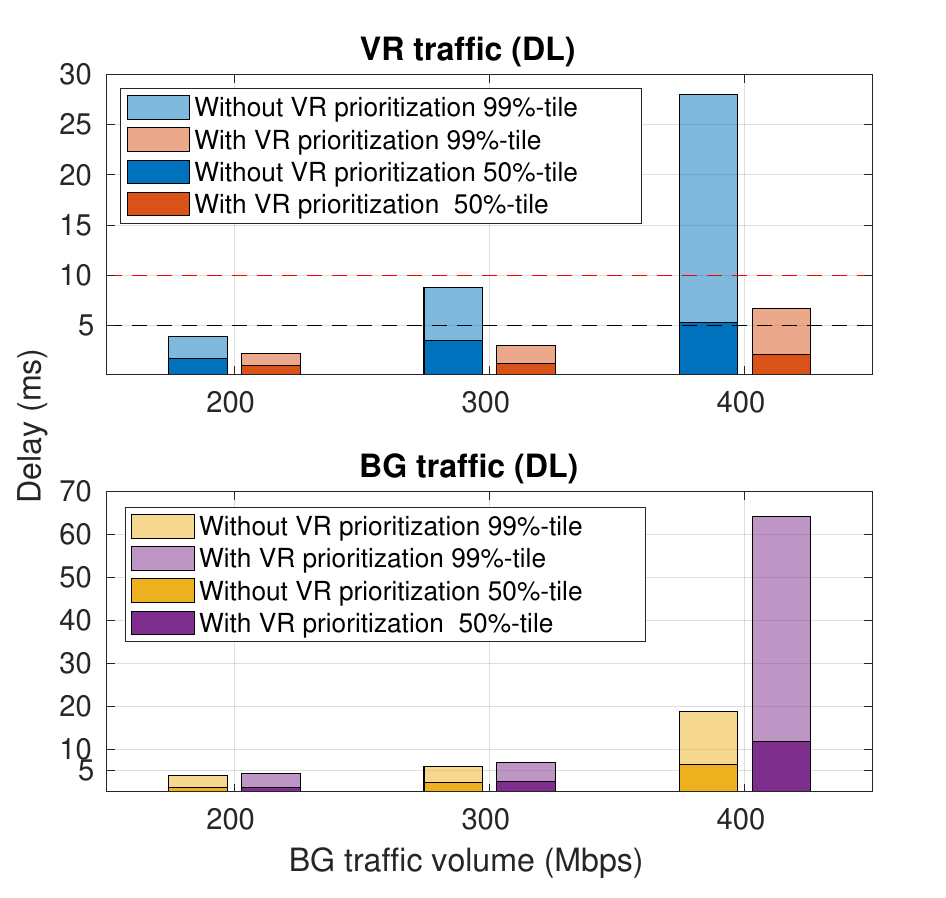}  
      \caption{A comparison of traffic packet delay for VR and BG traffic (DL) in both medium and worse delay scenarios, with and without VR prioritization.}
  \label{fig:figure6}
\end{figure}

Fig. \ref{fig:figure6} shows the median and worst-case packet delay (99th percentile) for both VR traffic and BG traffic with and without VR prioritization active in the DL. Blue and yellow bars show default First In First Out (FIFO) operation (i.e., no prioritization), and red and purple bars show the delay once VR prioritization is active. The 5 ms and 10 ms delay thresholds have been highlighted with dashed lines. VR requires very low latency, 5 ms or under would lead to the optimal experience, while delays above 10 ms would lead to a loss of video frames of Service (QoS) with giving priority to VR users in the AP and efficiently handle VR and an uneven experience that is noticeable to the player. Unprioritized VR traffic remains under 5 ms of worst-case delay for 200 Mbps, and 10 ms for 300 Mbps, maintaining a stable experience for the end user. At 400 Mbps however, the delay for VR traffic exceeds 25 ms, more than twice the maximum delay we need. Once VR prioritization is active, VR delay decreases in all cases, and for 400 Mbps we can observe a 76.27\% decrease, leading to a delay of less than 10 ms, and a smoother experience for the player. For BG traffic on the other hand, prioritization leads to an increase in the delay. This increase is negligible at lower loads, but for 400 Mbps the delay is more than doubled. Indeed, prioritizing one type of traffic will lead to worse delays for other traffic types. However, BG traffic could be a file download or video streaming through Youtube, which either do not have strict latency requirements (former) or can compensate through the use of buffering (latter), which VR traffic cannot use.

In this section, we have shown that Interactive Traffic Identification can be used to significantly improve the performance of VR traffic. By identifying the types of traffic being driven through an AP, the AP can then make smart decisions that prioritize delay-sensitive traffic at the expense of non-sensitive traffic. With our approach, we have achieved VR traffic delays 4.2x lower than without prioritization, with only a 2.3x higher delay in the BG traffic.

In particular, a VR flow contains all the traffic of a VR session. VR flows are divided in samples of duration $\omega$ from where features are extracted. Each sample is divided in $N$ sub-samples of duration ``$\tau$" used for the calculation of UL/DL cross-correlation. Finally, each sub-sample consists of a group of packets. VR flows (i.e., VR sessions) can last several minutes. However, the online traffic classification process is run only for the first sample of a VR flow. The total computation time for both feature extraction and classification of a sample is below one second by using a standard personal computer. A similar value can be obtained with an AP which may have dedicated resources to support ML operations. This value is very small compared to the VR session duration. Therefore, when the first sample of the VR flow is correctly classified as VR traffic, we can assume that the rest of the packets of the VR flow corresponds to VR traffic, without the necessity of running the classification process for the rest of the samples. Then, once the flow is identified as VR traffic, the prioritization process can be done without having an impact in terms of latency. Note that the training time -- considering training is done offline and before deploying the model -- is not evaluated since it is not relevant for real-time operations.
\newpage
\section{Conclusion and Future Work} \label{secconclusion}

Interactive XR/VR traffic identification over Wi-Fi can be helpful to decrease latency for XR users, improving network QoS and user experience. In this paper, we initially extracted statistical features from network characteristics across 15 settings, employing specific duration of sample (i.e. each of 1sec, 500ms, 200ms, 100ms, and 50ms) and a certain number of sub-samples (5, 10, and 20). We also included three features related to the correlation between DL and UL (i.e., RoNoP, RoTB and CC). Specifically, CC was computed based on a number of sub-samples (5, 10 and 20). Subsequently, in the classification phase, six ML methods were compared across all settings to determine the best-performing setting. This phase included feature selection (utilizing permutation importance) and hyperparameter tuning (via GridSearchCV). Evaluation metrics such as Accuracy, Precision, Recall, and F1-Score were employed for assessing the performance of the classification methods. The setting of $\omega = 500$ms and $N = 20$ emerged as the best-performing one, where two top-performing ML methods, DT and RF, achieved accuracies of 0.9874 and 0.9924, respectively.

To evaluate the interactive VR traffic identification model obtained from the training phase, two distinct processes were executed. Firstly, we evaluated the model by using (i) three datasets, each associated with a different user in a multi-user experimental setup, and (ii) a dataset collected when using SteamLink in single-user setup. These datasets were not part of the training dataset. In the multi-user dataset, using the top-performing classifiers (DT and RF) with the best setting ($\omega = 500$ms and $N = 20$), we achieved accuracies above 0.992 for all users, except for one instance where DT accuracy was 0.984. The total computational time for Client A was under 1 second per sample for both classifiers. In the single-user SteamLink dataset, both DT and RF also achieved high accuracy (0.984) with a computational time under 1 second per sample. The results indicate that our model performs well across these three streaming solutions. Therefore, it is likely that the classification model can work with other technologies, regardless of the source, as long as the traffic is encapsulated in transport protocols.

In general, our paper demonstrates that the ML-based traffic classification method offers a compelling balance between accuracy and computational efficiency. In our testing phase using the DT or RF classifier, we achieved an accuracy higher than 0.984 with a computational time under 1 second per sample. This indicates that the computational time is moderate, making it suitable for long-lasting real-time applications.

Secondly, we utilized a network simulator to investigate the potential impact of VR traffic identification in existing Wi-Fi networks. In a scenario involving an AP and two Stations (STAs), the first STA engaged in playing a VR game using ALVR, while the second STA received non-VR background traffic. The AP was configured to detect whether the traffic is VR or BG, and prioritize VR traffic, disregarding BG packets until all VR packets were transmitted. With this prioritization approach, we achieved VR traffic delays 4.2 times lower than without prioritization, accompanied by only a 2.3 times higher delay in the BG traffic.

As the future direction of our work, expanding the use of VR traces from more platforms/systems is important for capturing diverse scenarios not covered in the current dataset, allowing for re-training to generalize the model for new VR applications. Another important aspect is the consideration of further 'network effects', such as extra delays and packet losses, which can introduce variability in the received stream at the AP. Understanding and studying these network effects are essential for enhancing the reliability of VR applications. On the other hand, exploring new Wi-Fi mechanisms specifically designed for VR traffic is essential to improve the overall performance of VR experiences. This may include adjusting existing Wi-Fi standards to better accommodate the unique demands of VR applications.

\newpage

\section*{Acknowledgement}

The authors would like to thank Miguel Casasnovas for collecting the Unity packet traces. This work is partially funded by Wi-XR PID2021-123995NB-I00 (MCIU/AEI/FEDER,UE), MAX-R (101070072) EU, SGR 00955-2021 AGAUR, and by MCIN/AEI under the Maria de Maeztu Units of Excellence Programme (CEX2021-001195-M). This paper has also been partially funded by the Spanish Ministry of Science and Innovation MCIN/AEI/10.13039/501100011033 under ARTIST project (ref. PID2020- 115104RB-I00). 




\bibliographystyle{elsarticle-num} 
\bibliography{cas-refs}





\end{document}